\documentclass[aps,pre,reprint,superscriptaddress]{revtex4-1}
\usepackage{graphicx}
\usepackage{dcolumn}
\usepackage{subfigure}

\begin{document}

\title{Controllable two-dimensional robust multiferroic GaTeCl monolayer with giant ferroelectricity and superior multifunctions}

\author{Shi-Hao Zhang}
\affiliation{Beijing National Center for Condensed Matter Physics, Institute of Physics, Chinese Academy of Sciences, Beijing 100190, China}
\affiliation{School of Physical Sciences, University of Chinese Academy of Sciences, Beijing 100190, China}
\author{Bang-Gui Liu}
\email[]{bgliu@iphy.ac.cn}
\affiliation{Beijing National Center for Condensed Matter Physics, Institute of Physics, Chinese Academy of Sciences, Beijing 100190, China}
\affiliation{School of Physical Sciences, University of Chinese Academy of Sciences, Beijing 100190, China}

\date{\today}

\begin{abstract}
We propose through first-principles investigation that GaTeCl monolayer is an excellent two-dimensional multiferroics  with giant mechanical anisotropy and ferroelasticity. The calculated phonon spectrum, molecular dynamic simulations, and elastic modules confirm its stability, and our cleavage energy analysis shows that exfoliating one GaTeCl monolayer from the GaTeCl bulk is feasible. The calculated in-plane ferroelectric polarization reaches to 578 pC/m. The energy barriers per formula unit of the ferroelastic rotational and ferroelectric reversal transitions are 476.2 meV and 754.1 meV, respectively, which are the greatest in the two-dimensional multiferroics family so far, and makes the GaTeCl monolayer have robust ferroelasticity and ferroelectricity under room temperature. Uniaxial stress along the polarization direction can make the indirect gap transit to direct gap. Furthermore, the GaTeCl monolayer has giant piezoelectricity and optical second harmonic generation, especially for the range of visible light. These interesting mechanical, electronic, and optical properties of the GaTeCl monolayer show its great potential for multi-functional devices and nonlinear optoelectronic applications.
\end{abstract}


\maketitle


\section{Introduction}

Ferroelectric materials have played an important role in various electronic devices ~\cite{bune1998two,fong2004ferroelectricity,junquera2003critical,martin2016thin,RevModPhys.77.1083,damodaran2017three}. Thin-film ferroelectric materials have attracted a lot of attention for many applications, but in the ultrathin film, the small thickness induces enormous depolarizing field and suppresses ferroelectric dipoles perpendicular to the surface. For example, the polarization only exists when the thickness of film is larger than 24 \AA{} for BaTiO$_3$~\cite{fong2004ferroelectricity} or 12 \AA{} for PbTiO$_3$ film~\cite{junquera2003critical}.
In order to overcome this limitation, one sought ferroelectricity in atomistic monolayers, bilayers, and multi-layers, such as distorted 1T-MoS$_2$~\cite{shirodkar2014emergence}, AgBiP$_2$Se$_6$~\cite{xu2017monolayer}, group-IV monochalcogenides~\cite{fei2015giant,wu2016intrinsic,wang2017two,add1}, BiN~\cite{chen2017room}, unzipped graphene oxide monolayer~\cite{noor2016switchable}. Especially, the  group-IV monochalcogenides monolayer, as two-dimensional multiferroics~\cite{wu2016intrinsic,wang2017two,add1}, shows very interesting phenomena: the piezoelectric effect~\cite{fei2015giant}, giant photovoltaic effect~\cite{rangel2017large}, anisotropic thermoelectric response~\cite{medrano2016anisotropic}, and giant optical second harmonic generation~\cite{wang2017giant}. This monolayer can also be used to make new promising electrode materials for Li-ion batteries~\cite{zhou2016mx}  in energy storage and conversion technologies and can be used for water splitting~\cite{chowdhury2017monolayer} in catalysis technology.

On the other hand, there are only few two-dimensional multiferroics (e.g. ferroelasticity and ferroelectricity) so far. It is highly desirable to seek high-performance two-dimensional multiferroics. Because layered GaTeCl bulk material was synthesised experimentally in 1981~\cite{wilms1981gatecl}, it is of much interest to investigate the GaTeCl monolayer for promising ferroelectricity, ferroelasticity, and other useful properties. Through systematical first-principles investigation, we find that exfoliating one GaTeCl monolayer from the GaTeCl bulk is feasible, and the GaTeCl monolayer has strongly mechanical anisotropy, the giant ferroelasticity, and large ferroelectric polarization (578 pC/m). Furthermore, its great energy barriers of ferroelastic and ferroelectric transitions guarantee the robustness of ferroelasticity and ferroelectricity at high temperature. Our first-principles study also shows that under incident light, the GaTeCl monolayer has enormous optical second harmonic generation and the intensity is strongly anisotropic, which can be observed experimentally. More detailed results will be presented in the following.

\begin{figure*}[!htbp]
\includegraphics[width=0.95\textwidth]{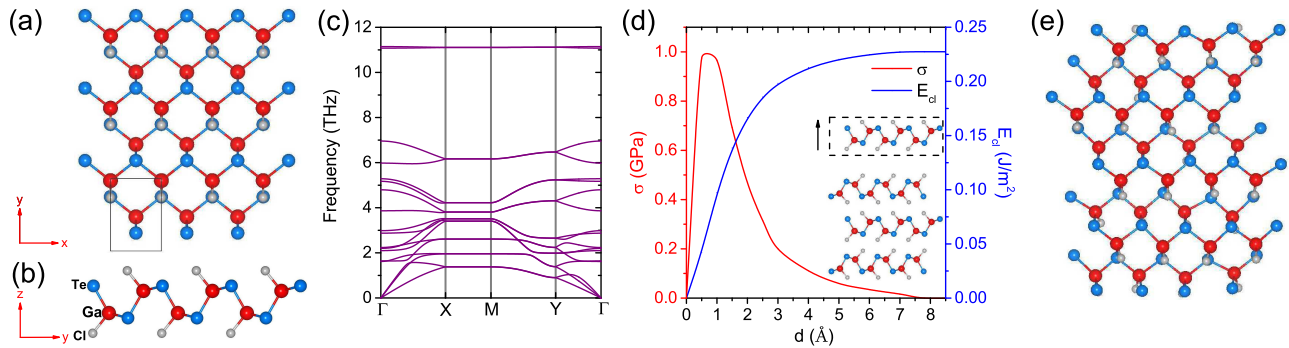}
\caption{~\label{fig1} The top view  (a) and side view (b) of the GaTeCl monolayer, where the red, blue, and gray balls represent Ga, Te, and Cl atoms, respectively. (c) The phonon spectra of GaTeCl monolayer, showing that there is no imaginary frequencies. (d) The cleavage strength $\sigma$ and cleavage energy $E_{cl}$ versus the separation distance $d$ in the process of exfoliating one GaTeCl monolayer from the GaTeCl bulk. (e) The top snapshot view of the GaTeCl monolayer after molecular dynamic simulations at the temperature of 550 K, showing the stability up to 550 K.}
\end{figure*}

\section{Computational methods}

The first-principles calculations are performed with the projector-augmented wave (PAW) method~\cite{PhysRevB.50.17953}  implemented in the Vienna Ab initio Simulation Package (VASP)~\cite{PhysRevB.47.558}. The generalized gradient approximation (GGA) by Perdew, Burke, and Ernzerhof~\cite{PhysRevLett.77.3865} is taken as the exchange-correlation potential. We carry out the Brillouin zone integration with a $\Gamma$-centered (15$\times$15$\times$1) Monkhorst-Pack grid~\cite{PhysRevB.13.5188}. The structures are fully optimized until all the Hellmann-Feynman forces on each atom are less than 0.02 eV/\AA{} and the total energy difference between two successive steps is smaller than $10^{-6}$ eV. Heyd-Scuseria-Ernzerhof (HSE06)~\cite{26,27,28} hybrid functional are also used in the energy band calculations. To ensure the structural stability of the monolayers, phonon spectra are calculated through the density functional perturbation theory in the PHONOPY program~\cite{TOGO20151}. Van der Waals correction with the D2 method~\cite{grimme2006semiempirical} is taken into account in the exfoliation energy calculation. We use Berry phase method to obtain the ferroelectric polarization ~\cite{king1993theory}. The minimum energy pathways of ferroelastic and ferroelectric transitions are calculated through nudged elastic band (NEB) method~\cite{mills1995reversible}. Optical second harmonic generation susceptibility tensor~\cite{sipe1993nonlinear,hughes1996calculation,sharma2003linear} is calculated with the ABINIT package~\cite{gonze2009abinit,gonze2016recent} with a dense k-point sampling of 60$\times$60$\times$1 and 70 electronic bands to ensure the convergence.

\section{results and discussion}

\subsection{Structure and stability}

\begin{figure}[!htbp]
\includegraphics[width=0.37\textwidth]{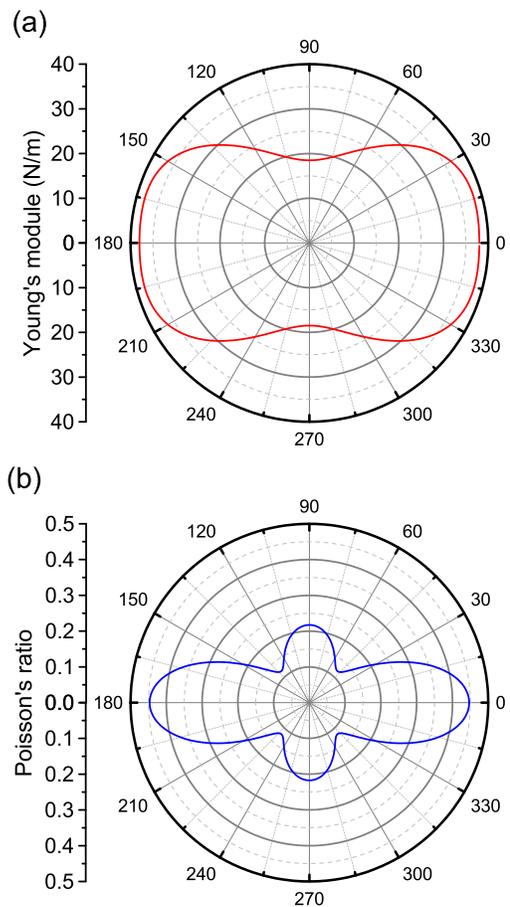}
\caption{~\label{fig2} The angular dependence of Young's module (a) and Poisson's ratio (b) of the GaTeCl monolayer. The zero degree corresponds to the x axis.}
\end{figure}

\begin{figure*}[!htbp]
\includegraphics[width=0.95\textwidth]{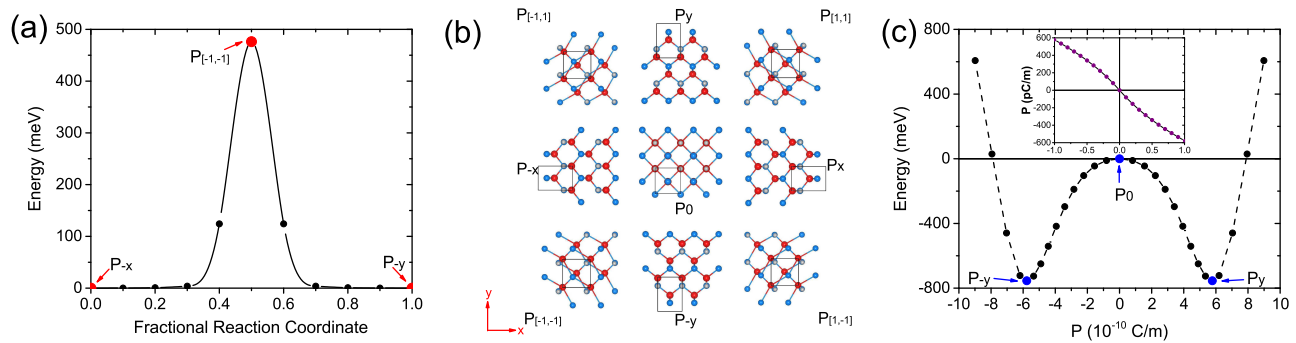}
\caption{~\label{fig3} (a) One of the minimal energy pathway of ferroelastic rotational transition in the GaTeCl monolayer. (b) The four degenerate ferroelastic and ferroelectric phases (P$_{\pm x}$ and P$_{\pm y}$), the four rotational transition states (P$_{[\pm1,\pm1]}$, ferroelectric), and the reversal transition state (P$_0$). (c) Double well potential of ferroelectric reversal transition in the GaTeCl monolayer, where the inset is calculated total polarization strength as a function of normalized displacement along the y axis.}
\end{figure*}

The optimized crystal structure of the GaTeCl monolayer can be seen in Fig. 1(a,b). The calculated lattice constants are 4.17 \AA{} along the x axis and 5.94 \AA{} along the y axis. Although the GaTeCl bulk obeys the Pnnm space group and is centrosymmetric, the GaTeCl monolayer obeys Pmn$2_1$ and becomes noncentrosymmetric. In the GaTeCl monolayer, each Ga atom connects three Te atoms and one Cl atom. The lengthes of Ga-Te and Ga-Cl bonds are 2.67 \AA{} and 2.21 \AA{}, and the angle the bonds make with respect to the y axis are 60.2$^{\text{o}}$ (Ga-Te) and 68.7$^{\text{o}}$ (Ga-Cl), respectively. The phonon spectra presented in Fig. 1(c) proves that there is no imaginary frequencies. To investigate the accessibility of exfoliation, we calculated the cleavage strength $\sigma$ and cleavage energy $E_{cl}$ shown in Fig. 1(d). Here, the cleavage strength is estimated by $\sigma = \partial E_{cl}/ \partial d $, and $d$ is the separation distance. The cleavage energy is 0.23 J/m$^{2}$, and the maximum cleavage strength is 1.0 GPa. They are comparable to the graphite (0.37 J/m$^{2}$ and 2.10 GPa)~\cite{wang2015measurement,Ca}. These confirm that exfoliation of the GaTeCl monolayer is feasible, and one GaTeCl monolayer can be exfoliated from the GaTeCl bulk~\cite{wilms1981gatecl}, as is shown in the inset of Fig. 1(d). The ab-initial molecular dynamic simulations of 4$\times$4 supercell (96 atoms) at 550 K with 3 ps time  show that the system is stable up to  550 K with small distortion of Cl atom's position, as shown in Fig. 1(e), which confirms the dynamical stability of the GaTeCl monolayer.

The calculated elastic constants, namely $C_{11} = 42.1 $ N/m, $C_{22} = 20.5 $ N/m, $C_{12} = 9.1 $ N/m, and $C_{66} = 13.8 $ N/m, satisfy the criteria of mechanical stability for two-dimensional materials: $C_{11}C_{22} > C_{12}C_{21}$ and $C_{66}>0$ ~\cite{32}. For the orthogonal symmetry, the Young's module and Poisson's ratio can be estimated by~\cite{wang2015electro}
\begin{eqnarray}
Y(\theta)=\frac{C_{11}C_{22}-C_{12}^2}{C_{22}\cos^4\theta+A \cos^2\theta \sin^2 \theta+C_{11}\sin^4\theta},
\end{eqnarray}
\begin{eqnarray}
\nu(\theta)=\frac{C_{12}\cos^4\theta-B \cos^2\theta \sin^2 \theta+C_{12}\sin^4\theta}{C_{22}\cos^4\theta+A \cos^2\theta \sin^2 \theta+C_{11}\sin^4\theta}.
\end{eqnarray}
where we have $A = (C_{11}C_{22}-C_{12}^2)/C_{66}-2C_{12}$ and $B = C_{11}+C_{22}-(C_{11}C_{22}-C_{12}^2)/C_{66}$, and $\theta$ is the angle of the direction with respect to the x axis. The calculated angular dependences of the  Young's module and Poisson's ratio are presented in Fig. 2. The Young's module varies from 20.5 to 42.1 N/m, and the Poisson's ratio ranges from 0.12 to 0.45 (one of the minima occurs at 48$^{\text{o}}$), which shows that the GaTeCl monolayer has giant mechanical anisotropy.

\begin{table*}[!htbp]
\caption{\label{table1}The semiconductor gap $E_g$, matrix elements $\eta _{11}$ and $\eta _{22}$ of transformation strain matrix $\eta_y$, polarization $P$, energy barriers of ferroelastic rotational transition $E_{bo}$ and ferroelectric reversal transition $E_{be}$, the changes of polarization strength under strain along polarization direction $e_{22}$, and maximum sheet second harmonic generation susceptibility $|\chi ^{(2)}|_{\rm max}$ of the GaTeCl monolayer and other two-dimensional multiferroics MX monolayers (M = Sn, Ge; X = S, Se)~\cite{fei2015giant,wu2016intrinsic,wang2017two,add1}.}
\begin{tabular}{p{1.5cm}<{\centering}p{1.5cm}<{\centering}p{1.5cm}<{\centering}p{1.5cm}<{\centering}p{1.8cm}<{\centering}p{1.8cm}<{\centering}p{1.8cm}
<{\centering}p{2.5cm}<{\centering}p{2.8cm}<{\centering}}
  \hline
  \hline
   Name & $E_g$ (eV) & $\eta _{11}$ & $\eta _{22}$ & $P$ (pC/m) & $E_{bo}$ (meV) & $E_{be}$ (meV) & $e_{22}$ (C/m) & $|\chi ^{(2)}|_{\rm max}$ (pm$^2$/V)\\ \hline
   SnS  & 1.37 & -0.019 & 0.028 & 247. & 8.6  & 33.1 & 18.1$\times 10^{-10}$ & 1.0$\times 10^6$ \\
   SnSe & 0.77 & -0.009 & 0.010 & 187. & 2.8  & 6.5  & 34.9$\times 10^{-10}$ & 4.0$\times 10^6$ \\
   GeS  & 1.23 & -0.070 & 0.141 & 441. & 56.4  & 464.0  & 4.6$\times 10^{-10}$ & 0.5$\times 10^6$ \\
   GeSe & 1.04 & -0.027 & 0.041 & 340. & 19.2 & 95.3  & 12.3$\times 10^{-10}$ & 5.2$\times 10^6$ \\
   GaTeCl & 2.31 & -0.116 & 0.279 & 578. & 476.2 & 754.1 & -2.17$\times 10^{-10}$ & 1.0$\times 10^6$ \\
   \hline
   \hline
\end{tabular}
\end{table*}

\begin{figure*}[!htbp]
\includegraphics[width=0.95\textwidth]{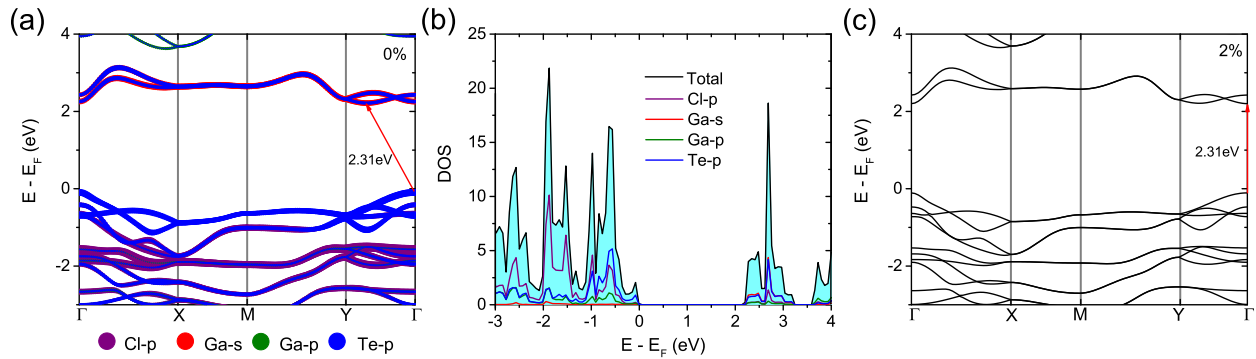}
\caption{~\label{fig4} (a) The energy bands of the GaTeCl monolayer without strain, where the weight contribution from Cl-p, Ga-s, Ga-p, and Te-p orbitals are also shown. (b) The corresponding total and partial density of states. (c) The energy bands of the GaTeCl monolayer under 2\% uniaxial strain along the y axis.}
\end{figure*}

\subsection{Ferroelasticity and Ferroelectricity}

It is noted that the lattice constants of the GaTeCl monolayer along the x and y axes are different. The system obeys mirror symmetry (M$_x$: x $\to$ -x) and the noncentrosymmetric cell undergoes the spontaneous strain along both of the x and y axes. Actually, there are four degenerate ground-state phases, P$_{\pm x}$ and P$_{\pm y}$, for the GaTeCl monolayer, as shown in Fig. 3. Here, we assign the x ferroelastic axis to P$_{\pm x}$, and the y ferroelastic axis to P$_{\pm y}$. For the each of P$_{\pm x}$ phases, there is a mirror symmetry M$_x$, and the monolayer expands along the y axis and is compressed along the x axis; and for the P$_{\pm y}$, x and y needs be interchanged. It can be seen by comparing with Fig. 1 that the ground-state phase presented above is P$_{-y}$ because its polarization is in the -y direction. A rotational transition path between P$_{-y}$ and P$_{-x}$ is shown in Fig. 3(a). This transition is through the para-elastic P$_{[-1,-1]}$ state. The transition can be described by a transformation strain matrix $\eta_y$ that can be obtained with Green-Lagrange strain tensor~\cite{li2016ferroelasticity}, $\eta _y = ([H^{-1}_{\rm ref}]^{T}H_{y}^{T}H_yH^{-1}_{\rm ref}-I)/2$, where $I$ is 2$\times$2 identity matrix [1,0;0,1], and $H_{y}=[4.17,0;0,5.94]$ and $H_{\rm ref}=[4.76,0;0,4.76]$ represent lattice parameters of the ferroelastic and paraelastic structures along the x and y axes, respectively. Then the diagonal matrix  $\eta _y$ for the P$_{\pm y}$ phases is [-0.116,0;0,0.279] which means that 27.9\% tensile strain along the y axis and 11.6\% compressive strain along the x axis. Similarly, the strain matrix $\eta_x$ for the P$_{\pm x}$ phase is [0.279,0;0,-0.116] implying 27.9\% tensile strain along the x axis and 11.6\% compressive strain along the y axis. The ferroelasticity of the GaTeCl monolayer is giant compared with other two-dimensional multiferroics (for instance, it is [-0.027,0;0,0.041] for GeSe monolayer)~\cite{wang2017two}, as shown in Table I. It is necessary to calculate the energy barrier in the ferroelastic transition. Nudged elastic band (NEB) calculation, as shown in Fig. 3(a), reveals that the energy barrier $E_{bo}$ is 476.2 meV per formula unit, which is much larger than those of other two-dimensional multiferroics, as shown in Table. ~\ref{table1}. This ensures the robustness of the ferroelasticity under room temperature.

Ferroelasticity does not always accompany ferroelectricity. For example, black phosphorus monolayer has ferroelasticity~\cite{mehboudi2016two}, but does not have ferroelectricity, because it has only P-P covalent bonds and  fails to form dipole moments. As for the GaTeCl monolayer, our Berry phase analysis of ferroelectric transition shows that the polarization $P$ is equivalent to 578 pC/m, as shown in Fig. 3(c) and Table~\ref{table1}. Compared with GeSe monolayer family~\cite{wu2016intrinsic,wang2017two}, the existence of Ga-Cl bond in the GaTeCl monolayer offers extra dipole moments and leads to the large polarization. To inspect the robustness of the ferroelectricity, we calculate the energy barrier per formula unit in the ferroelectric reversal transition over the paraelectric P$_0$ state, $E_{be}$, and it reaches to 754.1 meV, which implies that the ferroelectricity should survive beyond room temperature. The energy barrier of the GaTeCl monolayer is the greatest in the two-dimensional multiferroics family so far, as shown in Table~\ref{table1}. Our analysis shows that the Coulomb repulsive interaction between Te and Cl atoms makes the system electrostatically robust and increases the cost of phase transitions, thus the energy barriers of the ferroelastic and ferroelectric transitions are very high.

It is noted that each of the para-elastic P$_{[\pm1,\pm1]}$ states has ferroelectricity, and P$_0$ has neither ferroelasticity nor ferroelectricity. The energy barrier for the ferroelastic rotational transition over P$_{[\pm1,\pm1]}$ is much smaller than that of the ferroelectric reversal transition over P$_0$. It should be pointed out that the ferroelectric polarization can be successfully reversed through two steps of the rotational transitions with nearly half the energy barrier.

As usual, this multiferroics can have piezoelectric effect. The piezoelectric coefficients $e_{ijk}$ are third-rank tensors defined by $e_{ijk}=\partial P_i /\partial \epsilon _{jk}$~\cite{fei2015giant,gomes2015enhanced}, where $P_i$ is the polarization along the $i$ direction and $\epsilon _{jk}$ is the strain with the $jk$ subscripts. In this Pmn2$_1$ space group, there is only five non-zero piezoelectric constants in Voigt notation: $e_{21}$ = -1.00$\times10^{-10}$ C/m, $e_{22}$ = -2.17$\times10^{-10}$ C/m, $e_{23}$ = -0.18$\times10^{-10}$ C/m, $e_{16}$ = -1.64$\times10^{-10}$ C/m, and $e_{34}$ = 1.15$\times10^{-10}$ C/m. Especially, the $e_{22}$ (changes of polarization strength under strain along polarization direction) of the GaTeCl monolayer is negative while those of other piezoelectric materials (MX monolayer~\cite{fei2015giant}, transition metal dichalcogenide monolayer~\cite{duerloo2012intrinsic}, BN monolayer~\cite{duerloo2012intrinsic}, GaSe monolayer~\cite{li2015piezoelectricity} and others) are positive, as shown in Table I. This extraordinary feature can make the GaTeCl monolayer unique and applicable in tunable piezoelectric applications.

\subsection{Electronic structures}

Our electronic structure calculation shows that the GaTeCl monolayer is a semiconductor with indirect band gap of $E_g=2.31$ eV when the spin-orbit coupling is not taken into account. Its valence band maximum (VBM) is located at $\Gamma$ point, and the conduction band minimum (CBM) is between $\Gamma$ and $Y$ points, as shown in Fig. 4(a). The energy levels of VBM and CBM are at -8.76 eV and -6.36 eV with respect to the vacuum energy level, respectively. The effective mass at the VBM is 0.41$m_0$ along the x axis and 1.78$m_0$ along the y axis, and the effective mass at the CBM is 0.81$m_0$ along the y axis, where $m_0$ is the electron mass. The energy bands and density of states are presented in Fig. 4, where the spectrum weight contributions from Cl-p, Ga-s, Ga-p, and Te-p orbitals are also shown. The Cl-p, Ga-s, Ga-p, and Te-p orbitals contribute 17\%, 0\%, 12\%, and 71\% to the VBM, and 10\%, 46\%, 10\%, and 34\% to the CBM, respectively. When the spin-orbit coupling effect is taken into account, the GaTeCl monolayer is still a semiconductor with indirect band gap of 2.16 eV.

Uniaxial tensile stress along the y direction (the polarization direction) can tune the band gap of the GaTeCl monolayer. Without strain, the CBM of the monolayer is 39.1 meV lower than the lowest conduction band at the $\Gamma$ point, but 2\% tensile strain along the y direction (Poisson's effect is taken into consideration) makes the indirect band gap transit to the direct band gap, as shown in Fig. 5(c), and the original CBM remains 9.5 meV higher than the new CBM. Further tensile strain decreases the direct band gap, making the gap become 2.25 eV under 5\% uniaxial strain along the y direction. When the exchange-correlation functional is switched to HSE06, the semiconductor gap at zero strain becomes 2.82 eV, and the indirect-direct band gap transition due to uniaxial tensile strain remains true.

\subsection{Optical second harmonic generation}

When the incident light with electric field vector \textbf{E} irradiates the material, the material will be polarized and the electric polarization can be calculated by
\begin{eqnarray}
\textbf{P}=\textbf{P}_0+\chi ^{(1)}\textbf{E}+\chi ^{(2)}\textbf{E}^2+\chi ^{(3)}\textbf{E}^3+\cdots
\end{eqnarray}
where \textbf{P}$_0$ is the spontaneous polarization of the material in the absence of electric field. $\chi ^{(1)}$ is linear optical susceptibility and $\chi ^{(n)}$ is non-linear optical susceptibility ($n \ge 2$). For a monochromatic electric field, $E^a=E^a(\omega)e^{-i\omega t}+c.c.$, the second order polarization can be defined by
\begin{eqnarray}
P^{(2)}_a(t)=\chi ^{(2)}_{abc}(-2\omega ; \omega , \omega)E^b(\omega)E^c(\omega)e^{-2i\omega t}+c.c.
\end{eqnarray}
Because the GaTeCl monolayer obeys the point group C$_{2v}$ (mm2), it has five independent second harmonic generation susceptibility tensor elements: $\chi ^{(2)}_{yxx}$, $\chi ^{(2)}_{yyy}$, $\chi ^{(2)}_{yzz}$, $\chi ^{(2)}_{xyx}=\chi ^{(2)}_{xxy}$, and $\chi ^{(2)}_{zzy}=\chi ^{(2)}_{zyz}$. In order to compare with other materials, we use the sheet optical susceptibility $\chi _{\rm sheet}=\chi _{\rm bulk} \times L_z$~\cite{wang2017giant} where $L_z=2 \times 3.4+d_z$ includes the van der Waals thickness on both sides (3.4 \AA{} on each side) and the thickness of two-dimensional materials $d_z$ (2.73 \AA{} for the GaTeCl monolayer).

\begin{figure}[!htbp]
\includegraphics[width=0.38\textwidth]{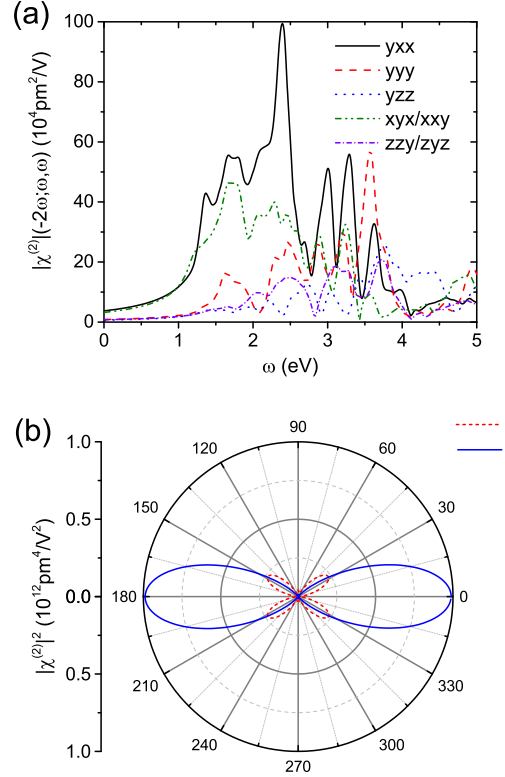}
\caption{~\label{fig5} (a) The sheet second harmonic generation susceptibilities in the GaTeCl monolayer. (b) Red (or blue) line represents the second harmonic generation intensity parallel (or perpendicular) to the polarization $E(\omega)$ of the incident electric field at $\omega$ = 2.39 eV. Here, $\theta$ is the angle made between $E(\omega)$ and the x axis.}
\end{figure}

The second harmonic generation susceptibility tensor elements as functions of photon energy are presented in Fig. 5(a).  The $\chi ^{(2)}_{yxx}$ component is most notable. As shown in Table I, the maximum of magnitude of the calculated sheet second harmonic generation susceptibility, $|\chi ^{(2)}|_{\rm max}$, reaches to 1.00$\times 10^6$ pm$^2$/V at 2.39 eV, which is comparable to those of MX monolayers (M = Sn, Ge; X = S, Se) and much larger than those of MoS$_{2}$ monolayer (3.02$\times 10^5$ pm$^2$/V) and BN monolayer (6.38 $\times 10^4$ pm$^2$/V)~\cite{wang2017giant}. This sheet second harmonic generation susceptibility is advantageous over the others for the range of visible light. The optical second harmonic generation effect can be observed by shedding linearly polarized laser beam on the material, and then measuring the different polarization component of outgoing response and inspecting the angular
dependence~\cite{PhysRevB.87.161403}. For the angle-dependent second harmonic generation susceptibilities, they can be computed by
\begin{eqnarray}
\chi ^{(2)}_{\Vert}(\theta)=(\chi ^{(2)}_{xyx}+\chi ^{(2)}_{yxx})\sin \theta \cos^2 \theta +\chi ^{(2)}_{yyy}\sin^3 \theta
\end{eqnarray}
\begin{eqnarray}
\chi ^{(2)}_{\bot}(\theta)=(-\chi ^{(2)}_{xyx}+\chi ^{(2)}_{yyy})\cos \theta \sin^2 \theta +\chi ^{(2)}_{yxx}\cos^3 \theta
\end{eqnarray}
where $\Vert$ ($\bot$) represent the polarization components of the second harmonic generation response parallel (perpendicular) to the polarization of incident electric field \textbf{E}. $\theta$ is the angle between the incident electric field \textbf{E} and the x axis. The second harmonic generation power, which can be measured in the experiment, is proportional to $|\chi ^{(2)}_{\Vert}(\theta)|^2$ and $|\chi ^{(2)}_{\bot}(\theta)|^2$. The results of $|\chi ^{(2)}_{\Vert}(\theta)|^2$ and $|\chi ^{(2)}_{\bot}(\theta)|^2$ at $\omega$ = 2.39 eV are presented in Fig. 5(b). The maximum of power ($|\chi ^{(2)}_{yxx}|^2$) occurs at $\theta$ = 0, reaching to 1.00$\times 10^{12}$ pm$^4$/V$^2$. The giant optical second harmonic generation makes the GaTeCl monolayer very interesting to nonlinear optoelectronic devices in the range of visible light.

\section{Conclusion}

We have predicted that the GaTeCl monolayer is a two-dimensional multiferroic semicondcutor with indirect bandgap of 2.31 eV. The low cleavage energy and strength indicate that the exfoliation of this monolayer from the GaTeCl bulk is feasible. Because of its special structure, the monolayer has strongly angle-dependent Young's module and Poisson's ratio. The GaTeCl monolayer has giant ferroelasticity and undergoes spontaneous 27.9\% tensile strain along the y axis and 11.6\% compressive strain along the x axis. Our NEB calculation shows that the energy barrier of ferroelastic rotational transition is 476.2 meV per formula unit which can make the ferroelasticity of the monolayer robust against distortion. Besides, the GaTeCl monolayer has giant ferroelectricity, with the polarization reaching to 578 pC/m. The energy barrier of 754.1 meV for the ferroelectric reversal transition ensures the strong robustness of the ferroelectricity under high temperature. Our piezoelectric calculation shows that the GaTeCl monolayer has very large piezoelectricity. The second harmonic generation susceptibility calculations reveal that the GaTeCl monolayer has giant optical second harmonic generation with the intensity being strongly anisotropic, advantageous for manipulation and applications for the visible light. Therefore, we believe that because of with all these excellent properties in one single two-dimensional monolayer material, the GaTeCl monolayer can be used to achieve promising multi-functional devices and nonlinear optoelectronic technologies.

\begin{acknowledgments}
This work is supported by the Nature Science Foundation of China (No.11574366), by the Strategic Priority Research Program of the Chinese Academy of Sciences (Grant No.XDB07000000), and by the Department of Science and Technology of China (Grant No.2016YFA0300701). The calculations were performed in the Milky Way \#2 supercomputer system at the National Supercomputer Center of Guangzhou, Guangzhou, China.
\end{acknowledgments}


\begin{thebibliography}{46}%
\makeatletter
\providecommand \@ifxundefined [1]{%
 \@ifx{#1\undefined}
}%
\providecommand \@ifnum [1]{%
 \ifnum #1\expandafter \@firstoftwo
 \else \expandafter \@secondoftwo
 \fi
}%
\providecommand \@ifx [1]{%
 \ifx #1\expandafter \@firstoftwo
 \else \expandafter \@secondoftwo
 \fi
}%
\providecommand \natexlab [1]{#1}%
\providecommand \enquote  [1]{``#1''}%
\providecommand \bibnamefont  [1]{#1}%
\providecommand \bibfnamefont [1]{#1}%
\providecommand \citenamefont [1]{#1}%
\providecommand \href@noop [0]{\@secondoftwo}%
\providecommand \href [0]{\begingroup \@sanitize@url \@href}%
\providecommand \@href[1]{\@@startlink{#1}\@@href}%
\providecommand \@@href[1]{\endgroup#1\@@endlink}%
\providecommand \@sanitize@url [0]{\catcode `\\12\catcode `\$12\catcode
  `\&12\catcode `\#12\catcode `\^12\catcode `\_12\catcode `\%12\relax}%
\providecommand \@@startlink[1]{}%
\providecommand \@@endlink[0]{}%
\providecommand \url  [0]{\begingroup\@sanitize@url \@url }%
\providecommand \@url [1]{\endgroup\@href {#1}{\urlprefix }}%
\providecommand \urlprefix  [0]{URL }%
\providecommand \Eprint [0]{\href }%
\providecommand \doibase [0]{http://dx.doi.org/}%
\providecommand \selectlanguage [0]{\@gobble}%
\providecommand \bibinfo  [0]{\@secondoftwo}%
\providecommand \bibfield  [0]{\@secondoftwo}%
\providecommand \translation [1]{[#1]}%
\providecommand \BibitemOpen [0]{}%
\providecommand \bibitemStop [0]{}%
\providecommand \bibitemNoStop [0]{.\EOS\space}%
\providecommand \EOS [0]{\spacefactor3000\relax}%
\providecommand \BibitemShut  [1]{\csname bibitem#1\endcsname}%
\let\auto@bib@innerbib\@empty
\bibitem [{\citenamefont {Bune}\ \emph {et~al.}(1998)\citenamefont {Bune},
  \citenamefont {Fridkin}, \citenamefont {Ducharme}, \citenamefont {Blinov}
  \emph {et~al.}}]{bune1998two}%
  \BibitemOpen
  \bibfield  {author} {\bibinfo {author} {\bibfnamefont {A.~V.}\ \bibnamefont
  {Bune}}, \bibinfo {author} {\bibfnamefont {V.~M.}\ \bibnamefont {Fridkin}},
  \bibinfo {author} {\bibfnamefont {S.}~\bibnamefont {Ducharme}}, \bibinfo
  {author} {\bibfnamefont {L.~M.}\ \bibnamefont {Blinov}},  \emph {et~al.},\
  }\href@noop {} {\bibfield  {journal} {\bibinfo  {journal} {Nature}\ }\textbf
  {\bibinfo {volume} {391}},\ \bibinfo {pages} {874} (\bibinfo {year}
  {1998})}\BibitemShut {NoStop}%
\bibitem [{\citenamefont {Fong}\ \emph {et~al.}(2004)\citenamefont {Fong},
  \citenamefont {Stephenson}, \citenamefont {Streiffer}, \citenamefont
  {Eastman}, \citenamefont {Auciello}, \citenamefont {Fuoss},\ and\
  \citenamefont {Thompson}}]{fong2004ferroelectricity}%
  \BibitemOpen
  \bibfield  {author} {\bibinfo {author} {\bibfnamefont {D.~D.}\ \bibnamefont
  {Fong}}, \bibinfo {author} {\bibfnamefont {G.~B.}\ \bibnamefont
  {Stephenson}}, \bibinfo {author} {\bibfnamefont {S.~K.}\ \bibnamefont
  {Streiffer}}, \bibinfo {author} {\bibfnamefont {J.~A.}\ \bibnamefont
  {Eastman}}, \bibinfo {author} {\bibfnamefont {O.}~\bibnamefont {Auciello}},
  \bibinfo {author} {\bibfnamefont {P.~H.}\ \bibnamefont {Fuoss}}, \ and\
  \bibinfo {author} {\bibfnamefont {C.}~\bibnamefont {Thompson}},\ }\href@noop
  {} {\bibfield  {journal} {\bibinfo  {journal} {Science}\ }\textbf {\bibinfo
  {volume} {304}},\ \bibinfo {pages} {1650} (\bibinfo {year}
  {2004})}\BibitemShut {NoStop}%
\bibitem [{\citenamefont {Junquera}\ and\ \citenamefont
  {Ghosez}(2003)}]{junquera2003critical}%
  \BibitemOpen
  \bibfield  {author} {\bibinfo {author} {\bibfnamefont {J.}~\bibnamefont
  {Junquera}}\ and\ \bibinfo {author} {\bibfnamefont {P.}~\bibnamefont
  {Ghosez}},\ }\href@noop {} {\bibfield  {journal} {\bibinfo  {journal}
  {Nature}\ }\textbf {\bibinfo {volume} {422}},\ \bibinfo {pages} {506}
  (\bibinfo {year} {2003})}\BibitemShut {NoStop}%
\bibitem [{\citenamefont {Martin}\ and\ \citenamefont
  {Rappe}(2016)}]{martin2016thin}%
  \BibitemOpen
  \bibfield  {author} {\bibinfo {author} {\bibfnamefont {L.~W.}\ \bibnamefont
  {Martin}}\ and\ \bibinfo {author} {\bibfnamefont {A.~M.}\ \bibnamefont
  {Rappe}},\ }\href@noop {} {\bibfield  {journal} {\bibinfo  {journal} {Nat.
  Rev. Mater.}\ }\textbf {\bibinfo {volume} {2}},\ \bibinfo {pages} {16087}
  (\bibinfo {year} {2016})}\BibitemShut {NoStop}%
\bibitem [{\citenamefont {Dawber}\ \emph {et~al.}(2005)\citenamefont {Dawber},
  \citenamefont {Rabe},\ and\ \citenamefont {Scott}}]{RevModPhys.77.1083}%
  \BibitemOpen
  \bibfield  {author} {\bibinfo {author} {\bibfnamefont {M.}~\bibnamefont
  {Dawber}}, \bibinfo {author} {\bibfnamefont {K.~M.}\ \bibnamefont {Rabe}}, \
  and\ \bibinfo {author} {\bibfnamefont {J.~F.}\ \bibnamefont {Scott}},\ }\href
  {\doibase 10.1103/RevModPhys.77.1083} {\bibfield  {journal} {\bibinfo
  {journal} {Rev. Mod. Phys.}\ }\textbf {\bibinfo {volume} {77}},\ \bibinfo
  {pages} {1083} (\bibinfo {year} {2005})}\BibitemShut {NoStop}%
\bibitem [{\citenamefont {Damodaran}\ \emph {et~al.}(2017)\citenamefont
  {Damodaran}, \citenamefont {Pandya}, \citenamefont {Agar}, \citenamefont
  {Cao}, \citenamefont {Vasudevan}, \citenamefont {Xu}, \citenamefont {Saremi},
  \citenamefont {Li}, \citenamefont {Kim}, \citenamefont {McCarter} \emph
  {et~al.}}]{damodaran2017three}%
  \BibitemOpen
  \bibfield  {author} {\bibinfo {author} {\bibfnamefont {A.~R.}\ \bibnamefont
  {Damodaran}}, \bibinfo {author} {\bibfnamefont {S.}~\bibnamefont {Pandya}},
  \bibinfo {author} {\bibfnamefont {J.~C.}\ \bibnamefont {Agar}}, \bibinfo
  {author} {\bibfnamefont {Y.}~\bibnamefont {Cao}}, \bibinfo {author}
  {\bibfnamefont {R.~K.}\ \bibnamefont {Vasudevan}}, \bibinfo {author}
  {\bibfnamefont {R.}~\bibnamefont {Xu}}, \bibinfo {author} {\bibfnamefont
  {S.}~\bibnamefont {Saremi}}, \bibinfo {author} {\bibfnamefont
  {Q.}~\bibnamefont {Li}}, \bibinfo {author} {\bibfnamefont {J.}~\bibnamefont
  {Kim}}, \bibinfo {author} {\bibfnamefont {M.~R.}\ \bibnamefont {McCarter}},
  \emph {et~al.},\ }\href@noop {} {\bibfield  {journal} {\bibinfo  {journal}
  {Adv. Mater.}\ } (\bibinfo {year} {2017})}\BibitemShut {NoStop}%
\bibitem [{\citenamefont {Shirodkar}\ and\ \citenamefont
  {Waghmare}(2014)}]{shirodkar2014emergence}%
  \BibitemOpen
  \bibfield  {author} {\bibinfo {author} {\bibfnamefont {S.~N.}\ \bibnamefont
  {Shirodkar}}\ and\ \bibinfo {author} {\bibfnamefont {U.~V.}\ \bibnamefont
  {Waghmare}},\ }\href@noop {} {\bibfield  {journal} {\bibinfo  {journal}
  {Phys. Rev. Lett.}\ }\textbf {\bibinfo {volume} {112}},\ \bibinfo {pages}
  {157601} (\bibinfo {year} {2014})}\BibitemShut {NoStop}%
\bibitem [{\citenamefont {Xu}\ \emph {et~al.}(2017)\citenamefont {Xu},
  \citenamefont {Xiang}, \citenamefont {Xia}, \citenamefont {Jiang},
  \citenamefont {Wan}, \citenamefont {He}, \citenamefont {Yin},\ and\
  \citenamefont {Liu}}]{xu2017monolayer}%
  \BibitemOpen
  \bibfield  {author} {\bibinfo {author} {\bibfnamefont {B.}~\bibnamefont
  {Xu}}, \bibinfo {author} {\bibfnamefont {H.}~\bibnamefont {Xiang}}, \bibinfo
  {author} {\bibfnamefont {Y.}~\bibnamefont {Xia}}, \bibinfo {author}
  {\bibfnamefont {K.}~\bibnamefont {Jiang}}, \bibinfo {author} {\bibfnamefont
  {X.}~\bibnamefont {Wan}}, \bibinfo {author} {\bibfnamefont {J.}~\bibnamefont
  {He}}, \bibinfo {author} {\bibfnamefont {J.}~\bibnamefont {Yin}}, \ and\
  \bibinfo {author} {\bibfnamefont {Z.}~\bibnamefont {Liu}},\ }\href@noop {}
  {\bibfield  {journal} {\bibinfo  {journal} {Nanoscale}\ }\textbf {\bibinfo
  {volume} {9}},\ \bibinfo {pages} {8427} (\bibinfo {year} {2017})}\BibitemShut
  {NoStop}%
\bibitem [{\citenamefont {Fei}\ \emph {et~al.}(2015)\citenamefont {Fei},
  \citenamefont {Li}, \citenamefont {Li},\ and\ \citenamefont
  {Yang}}]{fei2015giant}%
  \BibitemOpen
  \bibfield  {author} {\bibinfo {author} {\bibfnamefont {R.}~\bibnamefont
  {Fei}}, \bibinfo {author} {\bibfnamefont {W.}~\bibnamefont {Li}}, \bibinfo
  {author} {\bibfnamefont {J.}~\bibnamefont {Li}}, \ and\ \bibinfo {author}
  {\bibfnamefont {L.}~\bibnamefont {Yang}},\ }\href@noop {} {\bibfield
  {journal} {\bibinfo  {journal} {Appl. Phys. Lett.}\ }\textbf {\bibinfo
  {volume} {107}},\ \bibinfo {pages} {173104} (\bibinfo {year}
  {2015})}\BibitemShut {NoStop}%
\bibitem [{\citenamefont {Wu}\ and\ \citenamefont
  {Zeng}(2016)}]{wu2016intrinsic}%
  \BibitemOpen
  \bibfield  {author} {\bibinfo {author} {\bibfnamefont {M.}~\bibnamefont
  {Wu}}\ and\ \bibinfo {author} {\bibfnamefont {X.~C.}\ \bibnamefont {Zeng}},\
  }\href@noop {} {\bibfield  {journal} {\bibinfo  {journal} {Nano Lett.}\
  }\textbf {\bibinfo {volume} {16}},\ \bibinfo {pages} {3236} (\bibinfo {year}
  {2016})}\BibitemShut {NoStop}%
\bibitem [{\citenamefont {Wang}\ and\ \citenamefont
  {Qian}(2017{\natexlab{a}})}]{wang2017two}%
  \BibitemOpen
  \bibfield  {author} {\bibinfo {author} {\bibfnamefont {H.}~\bibnamefont
  {Wang}}\ and\ \bibinfo {author} {\bibfnamefont {X.}~\bibnamefont {Qian}},\
  }\href@noop {} {\bibfield  {journal} {\bibinfo  {journal} {2D Mater.}\
  }\textbf {\bibinfo {volume} {4}},\ \bibinfo {pages} {015042} (\bibinfo {year}
  {2017}{\natexlab{a}})}\BibitemShut {NoStop}%
\bibitem [{\citenamefont {Panday}\ and\ \citenamefont {Fregoso}(2017)}]{add1}%
  \BibitemOpen
  \bibfield  {author} {\bibinfo {author} {\bibfnamefont {S.~R.}\ \bibnamefont
  {Panday}}\ and\ \bibinfo {author} {\bibfnamefont {B.~M.}\ \bibnamefont
  {Fregoso}},\ }\href {\doibase 10.1088/1361-648X/aa8bfc} {\bibfield  {journal}
  {\bibinfo  {journal} {J. Phys.: Condens. Matter}\ }\textbf {\bibinfo {volume}
  {29}},\ \bibinfo {pages} {43LT01} (\bibinfo {year} {2017})}\BibitemShut
  {NoStop}%
\bibitem [{\citenamefont {Chen}\ \emph {et~al.}(2017)\citenamefont {Chen},
  \citenamefont {Zhang},\ and\ \citenamefont {Liu}}]{chen2017room}%
  \BibitemOpen
  \bibfield  {author} {\bibinfo {author} {\bibfnamefont {P.}~\bibnamefont
  {Chen}}, \bibinfo {author} {\bibfnamefont {X.-J.}\ \bibnamefont {Zhang}}, \
  and\ \bibinfo {author} {\bibfnamefont {B.-G.}\ \bibnamefont {Liu}},\
  }\href@noop {} {\bibfield  {journal} {\bibinfo  {journal} {arXiv preprint
  arXiv:1704.03424}\ } (\bibinfo {year} {2017})}\BibitemShut {NoStop}%
\bibitem [{\citenamefont {Noor-A-Alam}\ and\ \citenamefont
  {Shin}(2016)}]{noor2016switchable}%
  \BibitemOpen
  \bibfield  {author} {\bibinfo {author} {\bibfnamefont {M.}~\bibnamefont
  {Noor-A-Alam}}\ and\ \bibinfo {author} {\bibfnamefont {Y.-H.}\ \bibnamefont
  {Shin}},\ }\href@noop {} {\bibfield  {journal} {\bibinfo  {journal} {Phys.
  Chem. Chem. Phys.}\ }\textbf {\bibinfo {volume} {18}},\ \bibinfo {pages}
  {20443} (\bibinfo {year} {2016})}\BibitemShut {NoStop}%
\bibitem [{\citenamefont {Rangel}\ \emph {et~al.}(2017)\citenamefont {Rangel},
  \citenamefont {Fregoso}, \citenamefont {Mendoza}, \citenamefont {Morimoto},
  \citenamefont {Moore},\ and\ \citenamefont {Neaton}}]{rangel2017large}%
  \BibitemOpen
  \bibfield  {author} {\bibinfo {author} {\bibfnamefont {T.}~\bibnamefont
  {Rangel}}, \bibinfo {author} {\bibfnamefont {B.~M.}\ \bibnamefont {Fregoso}},
  \bibinfo {author} {\bibfnamefont {B.~S.}\ \bibnamefont {Mendoza}}, \bibinfo
  {author} {\bibfnamefont {T.}~\bibnamefont {Morimoto}}, \bibinfo {author}
  {\bibfnamefont {J.~E.}\ \bibnamefont {Moore}}, \ and\ \bibinfo {author}
  {\bibfnamefont {J.~B.}\ \bibnamefont {Neaton}},\ }\href@noop {} {\bibfield
  {journal} {\bibinfo  {journal} {Phys. Rev. Lett.}\ }\textbf {\bibinfo
  {volume} {119}},\ \bibinfo {pages} {067402} (\bibinfo {year}
  {2017})}\BibitemShut {NoStop}%
\bibitem [{\citenamefont {Medrano~Sandonas}\ \emph {et~al.}(2016)\citenamefont
  {Medrano~Sandonas}, \citenamefont {Teich}, \citenamefont {Gutierrez},
  \citenamefont {Lorenz}, \citenamefont {Pecchia}, \citenamefont {Seifert},\
  and\ \citenamefont {Cuniberti}}]{medrano2016anisotropic}%
  \BibitemOpen
  \bibfield  {author} {\bibinfo {author} {\bibfnamefont {L.}~\bibnamefont
  {Medrano~Sandonas}}, \bibinfo {author} {\bibfnamefont {D.}~\bibnamefont
  {Teich}}, \bibinfo {author} {\bibfnamefont {R.}~\bibnamefont {Gutierrez}},
  \bibinfo {author} {\bibfnamefont {T.}~\bibnamefont {Lorenz}}, \bibinfo
  {author} {\bibfnamefont {A.}~\bibnamefont {Pecchia}}, \bibinfo {author}
  {\bibfnamefont {G.}~\bibnamefont {Seifert}}, \ and\ \bibinfo {author}
  {\bibfnamefont {G.}~\bibnamefont {Cuniberti}},\ }\href@noop {} {\bibfield
  {journal} {\bibinfo  {journal} {J. Phys. Chem. C}\ }\textbf {\bibinfo
  {volume} {120}},\ \bibinfo {pages} {18841} (\bibinfo {year}
  {2016})}\BibitemShut {NoStop}%
\bibitem [{\citenamefont {Wang}\ and\ \citenamefont
  {Qian}(2017{\natexlab{b}})}]{wang2017giant}%
  \BibitemOpen
  \bibfield  {author} {\bibinfo {author} {\bibfnamefont {H.}~\bibnamefont
  {Wang}}\ and\ \bibinfo {author} {\bibfnamefont {X.}~\bibnamefont {Qian}},\
  }\href@noop {} {\bibfield  {journal} {\bibinfo  {journal} {Nano Lett.}\
  }\textbf {\bibinfo {volume} {17}},\ \bibinfo {pages} {5027} (\bibinfo {year}
  {2017}{\natexlab{b}})}\BibitemShut {NoStop}%
\bibitem [{\citenamefont {Zhou}(2016)}]{zhou2016mx}%
  \BibitemOpen
  \bibfield  {author} {\bibinfo {author} {\bibfnamefont {Y.}~\bibnamefont
  {Zhou}},\ }\href@noop {} {\bibfield  {journal} {\bibinfo  {journal} {J.
  Mater. Chem. A}\ }\textbf {\bibinfo {volume} {4}},\ \bibinfo {pages} {10906}
  (\bibinfo {year} {2016})}\BibitemShut {NoStop}%
\bibitem [{\citenamefont {Chowdhury}\ \emph {et~al.}(2017)\citenamefont
  {Chowdhury}, \citenamefont {Karmakar},\ and\ \citenamefont
  {Datta}}]{chowdhury2017monolayer}%
  \BibitemOpen
  \bibfield  {author} {\bibinfo {author} {\bibfnamefont {C.}~\bibnamefont
  {Chowdhury}}, \bibinfo {author} {\bibfnamefont {S.}~\bibnamefont {Karmakar}},
  \ and\ \bibinfo {author} {\bibfnamefont {A.}~\bibnamefont {Datta}},\
  }\href@noop {} {\bibfield  {journal} {\bibinfo  {journal} {J. Phys. Chem. C}\
  }\textbf {\bibinfo {volume} {121}},\ \bibinfo {pages} {7615} (\bibinfo {year}
  {2017})}\BibitemShut {NoStop}%
\bibitem [{\citenamefont {Wilms}\ and\ \citenamefont
  {Kniep}(1981)}]{wilms1981gatecl}%
  \BibitemOpen
  \bibfield  {author} {\bibinfo {author} {\bibfnamefont {A.}~\bibnamefont
  {Wilms}}\ and\ \bibinfo {author} {\bibfnamefont {R.}~\bibnamefont {Kniep}},\
  }\href@noop {} {\bibfield  {journal} {\bibinfo  {journal} {Z. Naturforsch.
  B}\ }\textbf {\bibinfo {volume} {36}},\ \bibinfo {pages} {1658} (\bibinfo
  {year} {1981})}\BibitemShut {NoStop}%
\bibitem [{\citenamefont {Bl\"ochl}(1994)}]{PhysRevB.50.17953}%
  \BibitemOpen
  \bibfield  {author} {\bibinfo {author} {\bibfnamefont {P.~E.}\ \bibnamefont
  {Bl\"ochl}},\ }\href {\doibase 10.1103/PhysRevB.50.17953} {\bibfield
  {journal} {\bibinfo  {journal} {Phys. Rev. B}\ }\textbf {\bibinfo {volume}
  {50}},\ \bibinfo {pages} {17953} (\bibinfo {year} {1994})}\BibitemShut
  {NoStop}%
\bibitem [{\citenamefont {Kresse}\ and\ \citenamefont
  {Hafner}(1993)}]{PhysRevB.47.558}%
  \BibitemOpen
  \bibfield  {author} {\bibinfo {author} {\bibfnamefont {G.}~\bibnamefont
  {Kresse}}\ and\ \bibinfo {author} {\bibfnamefont {J.}~\bibnamefont
  {Hafner}},\ }\href {\doibase 10.1103/PhysRevB.47.558} {\bibfield  {journal}
  {\bibinfo  {journal} {Phys. Rev. B}\ }\textbf {\bibinfo {volume} {47}},\
  \bibinfo {pages} {558} (\bibinfo {year} {1993})}\BibitemShut {NoStop}%
\bibitem [{\citenamefont {Perdew}\ \emph {et~al.}(1996)\citenamefont {Perdew},
  \citenamefont {Burke},\ and\ \citenamefont
  {Ernzerhof}}]{PhysRevLett.77.3865}%
  \BibitemOpen
  \bibfield  {author} {\bibinfo {author} {\bibfnamefont {J.~P.}\ \bibnamefont
  {Perdew}}, \bibinfo {author} {\bibfnamefont {K.}~\bibnamefont {Burke}}, \
  and\ \bibinfo {author} {\bibfnamefont {M.}~\bibnamefont {Ernzerhof}},\ }\href
  {\doibase 10.1103/PhysRevLett.77.3865} {\bibfield  {journal} {\bibinfo
  {journal} {Phys. Rev. Lett.}\ }\textbf {\bibinfo {volume} {77}},\ \bibinfo
  {pages} {3865} (\bibinfo {year} {1996})}\BibitemShut {NoStop}%
\bibitem [{\citenamefont {Monkhorst}\ and\ \citenamefont
  {Pack}(1976)}]{PhysRevB.13.5188}%
  \BibitemOpen
  \bibfield  {author} {\bibinfo {author} {\bibfnamefont {H.~J.}\ \bibnamefont
  {Monkhorst}}\ and\ \bibinfo {author} {\bibfnamefont {J.~D.}\ \bibnamefont
  {Pack}},\ }\href {\doibase 10.1103/PhysRevB.13.5188} {\bibfield  {journal}
  {\bibinfo  {journal} {Phys. Rev. B}\ }\textbf {\bibinfo {volume} {13}},\
  \bibinfo {pages} {5188} (\bibinfo {year} {1976})}\BibitemShut {NoStop}%
\bibitem [{\citenamefont {Heyd}\ \emph {et~al.}(2003)\citenamefont {Heyd},
  \citenamefont {Scuseria},\ and\ \citenamefont {Ernzerhof}}]{26}%
  \BibitemOpen
  \bibfield  {author} {\bibinfo {author} {\bibfnamefont {J.}~\bibnamefont
  {Heyd}}, \bibinfo {author} {\bibfnamefont {G.~E.}\ \bibnamefont {Scuseria}},
  \ and\ \bibinfo {author} {\bibfnamefont {M.}~\bibnamefont {Ernzerhof}},\
  }\href {\doibase 10.1063/1.1564060} {\bibfield  {journal} {\bibinfo
  {journal} {J. Chem. Phys.}\ }\textbf {\bibinfo {volume} {118}},\ \bibinfo
  {pages} {8207} (\bibinfo {year} {2003})}\BibitemShut {NoStop}%
\bibitem [{\citenamefont {Krukau}\ \emph {et~al.}(2006)\citenamefont {Krukau},
  \citenamefont {Vydrov}, \citenamefont {Izmaylov},\ and\ \citenamefont
  {Scuseria}}]{27}%
  \BibitemOpen
  \bibfield  {author} {\bibinfo {author} {\bibfnamefont {A.~V.}\ \bibnamefont
  {Krukau}}, \bibinfo {author} {\bibfnamefont {O.~A.}\ \bibnamefont {Vydrov}},
  \bibinfo {author} {\bibfnamefont {A.~F.}\ \bibnamefont {Izmaylov}}, \ and\
  \bibinfo {author} {\bibfnamefont {G.~E.}\ \bibnamefont {Scuseria}},\ }\href
  {\doibase 10.1063/1.2404663} {\bibfield  {journal} {\bibinfo  {journal} {J.
  Chem. Phys.}\ }\textbf {\bibinfo {volume} {125}},\ \bibinfo {pages} {224106}
  (\bibinfo {year} {2006})}\BibitemShut {NoStop}%
\bibitem [{\citenamefont {Mori-S\'anchez}\ \emph {et~al.}(2008)\citenamefont
  {Mori-S\'anchez}, \citenamefont {Cohen},\ and\ \citenamefont {Yang}}]{28}%
  \BibitemOpen
  \bibfield  {author} {\bibinfo {author} {\bibfnamefont {P.}~\bibnamefont
  {Mori-S\'anchez}}, \bibinfo {author} {\bibfnamefont {A.~J.}\ \bibnamefont
  {Cohen}}, \ and\ \bibinfo {author} {\bibfnamefont {W.}~\bibnamefont {Yang}},\
  }\href {\doibase 10.1103/PhysRevLett.100.146401} {\bibfield  {journal}
  {\bibinfo  {journal} {Phys. Rev. Lett.}\ }\textbf {\bibinfo {volume} {100}},\
  \bibinfo {pages} {146401} (\bibinfo {year} {2008})}\BibitemShut {NoStop}%
\bibitem [{\citenamefont {Togo}\ and\ \citenamefont
  {Tanaka}(2015)}]{TOGO20151}%
  \BibitemOpen
  \bibfield  {author} {\bibinfo {author} {\bibfnamefont {A.}~\bibnamefont
  {Togo}}\ and\ \bibinfo {author} {\bibfnamefont {I.}~\bibnamefont {Tanaka}},\
  }\href {\doibase http://dx.doi.org/10.1016/j.scriptamat.2015.07.021}
  {\bibfield  {journal} {\bibinfo  {journal} {Scr. Mater.}\ }\textbf {\bibinfo
  {volume} {108}},\ \bibinfo {pages} {1 } (\bibinfo {year} {2015})}\BibitemShut
  {NoStop}%
\bibitem [{\citenamefont {Grimme}(2006)}]{grimme2006semiempirical}%
  \BibitemOpen
  \bibfield  {author} {\bibinfo {author} {\bibfnamefont {S.}~\bibnamefont
  {Grimme}},\ }\href@noop {} {\bibfield  {journal} {\bibinfo  {journal} {J.
  Comput. Chem.}\ }\textbf {\bibinfo {volume} {27}},\ \bibinfo {pages} {1787}
  (\bibinfo {year} {2006})}\BibitemShut {NoStop}%
\bibitem [{\citenamefont {King-Smith}\ and\ \citenamefont
  {Vanderbilt}(1993)}]{king1993theory}%
  \BibitemOpen
  \bibfield  {author} {\bibinfo {author} {\bibfnamefont {R.}~\bibnamefont
  {King-Smith}}\ and\ \bibinfo {author} {\bibfnamefont {D.}~\bibnamefont
  {Vanderbilt}},\ }\href@noop {} {\bibfield  {journal} {\bibinfo  {journal}
  {Phys. Rev. B}\ }\textbf {\bibinfo {volume} {47}},\ \bibinfo {pages} {1651}
  (\bibinfo {year} {1993})}\BibitemShut {NoStop}%
\bibitem [{\citenamefont {Mills}\ \emph {et~al.}(1995)\citenamefont {Mills},
  \citenamefont {J{\'o}nsson},\ and\ \citenamefont
  {Schenter}}]{mills1995reversible}%
  \BibitemOpen
  \bibfield  {author} {\bibinfo {author} {\bibfnamefont {G.}~\bibnamefont
  {Mills}}, \bibinfo {author} {\bibfnamefont {H.}~\bibnamefont {J{\'o}nsson}},
  \ and\ \bibinfo {author} {\bibfnamefont {G.~K.}\ \bibnamefont {Schenter}},\
  }\href@noop {} {\bibfield  {journal} {\bibinfo  {journal} {Surf. Sci.}\
  }\textbf {\bibinfo {volume} {324}},\ \bibinfo {pages} {305} (\bibinfo {year}
  {1995})}\BibitemShut {NoStop}%
\bibitem [{\citenamefont {Sipe}\ and\ \citenamefont
  {Ghahramani}(1993)}]{sipe1993nonlinear}%
  \BibitemOpen
  \bibfield  {author} {\bibinfo {author} {\bibfnamefont {J.}~\bibnamefont
  {Sipe}}\ and\ \bibinfo {author} {\bibfnamefont {E.}~\bibnamefont
  {Ghahramani}},\ }\href@noop {} {\bibfield  {journal} {\bibinfo  {journal}
  {Phys. Rev. B}\ }\textbf {\bibinfo {volume} {48}},\ \bibinfo {pages} {11705}
  (\bibinfo {year} {1993})}\BibitemShut {NoStop}%
\bibitem [{\citenamefont {Hughes}\ and\ \citenamefont
  {Sipe}(1996)}]{hughes1996calculation}%
  \BibitemOpen
  \bibfield  {author} {\bibinfo {author} {\bibfnamefont {J.~L.}\ \bibnamefont
  {Hughes}}\ and\ \bibinfo {author} {\bibfnamefont {J.}~\bibnamefont {Sipe}},\
  }\href@noop {} {\bibfield  {journal} {\bibinfo  {journal} {Phys. Rev. B}\
  }\textbf {\bibinfo {volume} {53}},\ \bibinfo {pages} {10751} (\bibinfo {year}
  {1996})}\BibitemShut {NoStop}%
\bibitem [{\citenamefont {Sharma}\ \emph {et~al.}(2003)\citenamefont {Sharma},
  \citenamefont {Dewhurst},\ and\ \citenamefont
  {Ambrosch-Draxl}}]{sharma2003linear}%
  \BibitemOpen
  \bibfield  {author} {\bibinfo {author} {\bibfnamefont {S.}~\bibnamefont
  {Sharma}}, \bibinfo {author} {\bibfnamefont {J.}~\bibnamefont {Dewhurst}}, \
  and\ \bibinfo {author} {\bibfnamefont {C.}~\bibnamefont {Ambrosch-Draxl}},\
  }\href@noop {} {\bibfield  {journal} {\bibinfo  {journal} {Phys. Rev. B}\
  }\textbf {\bibinfo {volume} {67}},\ \bibinfo {pages} {165332} (\bibinfo
  {year} {2003})}\BibitemShut {NoStop}%
\bibitem [{\citenamefont {Gonze}\ \emph {et~al.}(2009)\citenamefont {Gonze},
  \citenamefont {Amadon}, \citenamefont {Anglade}, \citenamefont {Beuken},
  \citenamefont {Bottin}, \citenamefont {Boulanger}, \citenamefont {Bruneval},
  \citenamefont {Caliste}, \citenamefont {Caracas}, \citenamefont
  {C{\^o}t{\'e}} \emph {et~al.}}]{gonze2009abinit}%
  \BibitemOpen
  \bibfield  {author} {\bibinfo {author} {\bibfnamefont {X.}~\bibnamefont
  {Gonze}}, \bibinfo {author} {\bibfnamefont {B.}~\bibnamefont {Amadon}},
  \bibinfo {author} {\bibfnamefont {P.-M.}\ \bibnamefont {Anglade}}, \bibinfo
  {author} {\bibfnamefont {J.-M.}\ \bibnamefont {Beuken}}, \bibinfo {author}
  {\bibfnamefont {F.}~\bibnamefont {Bottin}}, \bibinfo {author} {\bibfnamefont
  {P.}~\bibnamefont {Boulanger}}, \bibinfo {author} {\bibfnamefont
  {F.}~\bibnamefont {Bruneval}}, \bibinfo {author} {\bibfnamefont
  {D.}~\bibnamefont {Caliste}}, \bibinfo {author} {\bibfnamefont
  {R.}~\bibnamefont {Caracas}}, \bibinfo {author} {\bibfnamefont
  {M.}~\bibnamefont {C{\^o}t{\'e}}},  \emph {et~al.},\ }\href@noop {}
  {\bibfield  {journal} {\bibinfo  {journal} {Comput. Phys. Commun.}\ }\textbf
  {\bibinfo {volume} {180}},\ \bibinfo {pages} {2582} (\bibinfo {year}
  {2009})}\BibitemShut {NoStop}%
\bibitem [{\citenamefont {Gonze}\ \emph {et~al.}(2016)\citenamefont {Gonze},
  \citenamefont {Jollet}, \citenamefont {Araujo}, \citenamefont {Adams},
  \citenamefont {Amadon}, \citenamefont {Applencourt}, \citenamefont {Audouze},
  \citenamefont {Beuken}, \citenamefont {Bieder}, \citenamefont {Bokhanchuk}
  \emph {et~al.}}]{gonze2016recent}%
  \BibitemOpen
  \bibfield  {author} {\bibinfo {author} {\bibfnamefont {X.}~\bibnamefont
  {Gonze}}, \bibinfo {author} {\bibfnamefont {F.}~\bibnamefont {Jollet}},
  \bibinfo {author} {\bibfnamefont {F.~A.}\ \bibnamefont {Araujo}}, \bibinfo
  {author} {\bibfnamefont {D.}~\bibnamefont {Adams}}, \bibinfo {author}
  {\bibfnamefont {B.}~\bibnamefont {Amadon}}, \bibinfo {author} {\bibfnamefont
  {T.}~\bibnamefont {Applencourt}}, \bibinfo {author} {\bibfnamefont
  {C.}~\bibnamefont {Audouze}}, \bibinfo {author} {\bibfnamefont {J.-M.}\
  \bibnamefont {Beuken}}, \bibinfo {author} {\bibfnamefont {J.}~\bibnamefont
  {Bieder}}, \bibinfo {author} {\bibfnamefont {A.}~\bibnamefont {Bokhanchuk}},
  \emph {et~al.},\ }\href@noop {} {\bibfield  {journal} {\bibinfo  {journal}
  {Comput. Phys. Commun.}\ }\textbf {\bibinfo {volume} {205}},\ \bibinfo
  {pages} {106} (\bibinfo {year} {2016})}\BibitemShut {NoStop}%
\bibitem [{\citenamefont {Wang}\ \emph
  {et~al.}(2015{\natexlab{a}})\citenamefont {Wang}, \citenamefont {Dai},
  \citenamefont {Li}, \citenamefont {Yang}, \citenamefont {Srolovitz},\ and\
  \citenamefont {Zheng}}]{wang2015measurement}%
  \BibitemOpen
  \bibfield  {author} {\bibinfo {author} {\bibfnamefont {W.}~\bibnamefont
  {Wang}}, \bibinfo {author} {\bibfnamefont {S.}~\bibnamefont {Dai}}, \bibinfo
  {author} {\bibfnamefont {X.}~\bibnamefont {Li}}, \bibinfo {author}
  {\bibfnamefont {J.}~\bibnamefont {Yang}}, \bibinfo {author} {\bibfnamefont
  {D.~J.}\ \bibnamefont {Srolovitz}}, \ and\ \bibinfo {author} {\bibfnamefont
  {Q.}~\bibnamefont {Zheng}},\ }\href {\doibase 10.1038/ncomms8853} {\bibfield
  {journal} {\bibinfo  {journal} {Nat. Commun.}\ }\textbf {\bibinfo {volume}
  {6}},\ \bibinfo {pages} {7853} (\bibinfo {year}
  {2015}{\natexlab{a}})}\BibitemShut {NoStop}%
\bibitem [{\citenamefont {Zhao}\ \emph {et~al.}(2014)\citenamefont {Zhao},
  \citenamefont {Li},\ and\ \citenamefont {Yang}}]{Ca}%
  \BibitemOpen
  \bibfield  {author} {\bibinfo {author} {\bibfnamefont {S.}~\bibnamefont
  {Zhao}}, \bibinfo {author} {\bibfnamefont {Z.}~\bibnamefont {Li}}, \ and\
  \bibinfo {author} {\bibfnamefont {J.}~\bibnamefont {Yang}},\ }\href {\doibase
  10.1021/ja5065125} {\bibfield  {journal} {\bibinfo  {journal} {J. Am. Chem.
  Soc.}\ }\textbf {\bibinfo {volume} {136}},\ \bibinfo {pages} {13313}
  (\bibinfo {year} {2014})}\BibitemShut {NoStop}%
\bibitem [{\citenamefont {Andrew}\ \emph {et~al.}(2012)\citenamefont {Andrew},
  \citenamefont {Mapasha}, \citenamefont {Ukpong},\ and\ \citenamefont
  {Chetty}}]{32}%
  \BibitemOpen
  \bibfield  {author} {\bibinfo {author} {\bibfnamefont {R.~C.}\ \bibnamefont
  {Andrew}}, \bibinfo {author} {\bibfnamefont {R.~E.}\ \bibnamefont {Mapasha}},
  \bibinfo {author} {\bibfnamefont {A.~M.}\ \bibnamefont {Ukpong}}, \ and\
  \bibinfo {author} {\bibfnamefont {N.}~\bibnamefont {Chetty}},\ }\href
  {\doibase 10.1103/PhysRevB.85.125428} {\bibfield  {journal} {\bibinfo
  {journal} {Phys. Rev. B}\ }\textbf {\bibinfo {volume} {85}},\ \bibinfo
  {pages} {125428} (\bibinfo {year} {2012})}\BibitemShut {NoStop}%
\bibitem [{\citenamefont {Wang}\ \emph
  {et~al.}(2015{\natexlab{b}})\citenamefont {Wang}, \citenamefont {Kutana},
  \citenamefont {Zou},\ and\ \citenamefont {Yakobson}}]{wang2015electro}%
  \BibitemOpen
  \bibfield  {author} {\bibinfo {author} {\bibfnamefont {L.}~\bibnamefont
  {Wang}}, \bibinfo {author} {\bibfnamefont {A.}~\bibnamefont {Kutana}},
  \bibinfo {author} {\bibfnamefont {X.}~\bibnamefont {Zou}}, \ and\ \bibinfo
  {author} {\bibfnamefont {B.~I.}\ \bibnamefont {Yakobson}},\ }\href@noop {}
  {\bibfield  {journal} {\bibinfo  {journal} {Nanoscale}\ }\textbf {\bibinfo
  {volume} {7}},\ \bibinfo {pages} {9746} (\bibinfo {year}
  {2015}{\natexlab{b}})}\BibitemShut {NoStop}%
\bibitem [{\citenamefont {Li}\ and\ \citenamefont
  {Li}(2016)}]{li2016ferroelasticity}%
  \BibitemOpen
  \bibfield  {author} {\bibinfo {author} {\bibfnamefont {W.}~\bibnamefont
  {Li}}\ and\ \bibinfo {author} {\bibfnamefont {J.}~\bibnamefont {Li}},\
  }\href@noop {} {\bibfield  {journal} {\bibinfo  {journal} {Nat. Commun.}\
  }\textbf {\bibinfo {volume} {7}},\ \bibinfo {pages} {10843} (\bibinfo {year}
  {2016})}\BibitemShut {NoStop}%
\bibitem [{\citenamefont {Mehboudi}\ \emph {et~al.}(2016)\citenamefont
  {Mehboudi}, \citenamefont {Dorio}, \citenamefont {Zhu}, \citenamefont
  {van~der Zande}, \citenamefont {Churchill}, \citenamefont {Pacheco-Sanjuan},
  \citenamefont {Harriss}, \citenamefont {Kumar},\ and\ \citenamefont
  {Barraza-Lopez}}]{mehboudi2016two}%
  \BibitemOpen
  \bibfield  {author} {\bibinfo {author} {\bibfnamefont {M.}~\bibnamefont
  {Mehboudi}}, \bibinfo {author} {\bibfnamefont {A.~M.}\ \bibnamefont {Dorio}},
  \bibinfo {author} {\bibfnamefont {W.}~\bibnamefont {Zhu}}, \bibinfo {author}
  {\bibfnamefont {A.}~\bibnamefont {van~der Zande}}, \bibinfo {author}
  {\bibfnamefont {H.~O.}\ \bibnamefont {Churchill}}, \bibinfo {author}
  {\bibfnamefont {A.~A.}\ \bibnamefont {Pacheco-Sanjuan}}, \bibinfo {author}
  {\bibfnamefont {E.~O.}\ \bibnamefont {Harriss}}, \bibinfo {author}
  {\bibfnamefont {P.}~\bibnamefont {Kumar}}, \ and\ \bibinfo {author}
  {\bibfnamefont {S.}~\bibnamefont {Barraza-Lopez}},\ }\href@noop {} {\bibfield
   {journal} {\bibinfo  {journal} {Nano Lett.}\ }\textbf {\bibinfo {volume}
  {16}},\ \bibinfo {pages} {1704} (\bibinfo {year} {2016})}\BibitemShut
  {NoStop}%
\bibitem [{\citenamefont {Gomes}\ \emph {et~al.}(2015)\citenamefont {Gomes},
  \citenamefont {Carvalho},\ and\ \citenamefont {Neto}}]{gomes2015enhanced}%
  \BibitemOpen
  \bibfield  {author} {\bibinfo {author} {\bibfnamefont {L.~C.}\ \bibnamefont
  {Gomes}}, \bibinfo {author} {\bibfnamefont {A.}~\bibnamefont {Carvalho}}, \
  and\ \bibinfo {author} {\bibfnamefont {A.~C.}\ \bibnamefont {Neto}},\
  }\href@noop {} {\bibfield  {journal} {\bibinfo  {journal} {Phys. Rev. B}\
  }\textbf {\bibinfo {volume} {92}},\ \bibinfo {pages} {214103} (\bibinfo
  {year} {2015})}\BibitemShut {NoStop}%
\bibitem [{\citenamefont {Duerloo}\ \emph {et~al.}(2012)\citenamefont
  {Duerloo}, \citenamefont {Ong},\ and\ \citenamefont
  {Reed}}]{duerloo2012intrinsic}%
  \BibitemOpen
  \bibfield  {author} {\bibinfo {author} {\bibfnamefont {K.-A.~N.}\
  \bibnamefont {Duerloo}}, \bibinfo {author} {\bibfnamefont {M.~T.}\
  \bibnamefont {Ong}}, \ and\ \bibinfo {author} {\bibfnamefont {E.~J.}\
  \bibnamefont {Reed}},\ }\href@noop {} {\bibfield  {journal} {\bibinfo
  {journal} {J. Phys. Chem. Lett.}\ }\textbf {\bibinfo {volume} {3}},\ \bibinfo
  {pages} {2871} (\bibinfo {year} {2012})}\BibitemShut {NoStop}%
\bibitem [{\citenamefont {Li}\ and\ \citenamefont
  {Li}(2015)}]{li2015piezoelectricity}%
  \BibitemOpen
  \bibfield  {author} {\bibinfo {author} {\bibfnamefont {W.}~\bibnamefont
  {Li}}\ and\ \bibinfo {author} {\bibfnamefont {J.}~\bibnamefont {Li}},\
  }\href@noop {} {\bibfield  {journal} {\bibinfo  {journal} {Nano Res.}\
  }\textbf {\bibinfo {volume} {8}},\ \bibinfo {pages} {3796} (\bibinfo {year}
  {2015})}\BibitemShut {NoStop}%
\bibitem [{\citenamefont {Kumar}\ \emph {et~al.}(2013)\citenamefont {Kumar},
  \citenamefont {Najmaei}, \citenamefont {Cui}, \citenamefont {Ceballos},
  \citenamefont {Ajayan}, \citenamefont {Lou},\ and\ \citenamefont
  {Zhao}}]{PhysRevB.87.161403}%
  \BibitemOpen
  \bibfield  {author} {\bibinfo {author} {\bibfnamefont {N.}~\bibnamefont
  {Kumar}}, \bibinfo {author} {\bibfnamefont {S.}~\bibnamefont {Najmaei}},
  \bibinfo {author} {\bibfnamefont {Q.}~\bibnamefont {Cui}}, \bibinfo {author}
  {\bibfnamefont {F.}~\bibnamefont {Ceballos}}, \bibinfo {author}
  {\bibfnamefont {P.~M.}\ \bibnamefont {Ajayan}}, \bibinfo {author}
  {\bibfnamefont {J.}~\bibnamefont {Lou}}, \ and\ \bibinfo {author}
  {\bibfnamefont {H.}~\bibnamefont {Zhao}},\ }\href {\doibase
  10.1103/PhysRevB.87.161403} {\bibfield  {journal} {\bibinfo  {journal} {Phys.
  Rev. B}\ }\textbf {\bibinfo {volume} {87}},\ \bibinfo {pages} {161403}
  (\bibinfo {year} {2013})}\BibitemShut {NoStop}%
\end{thebibliography}
%

\end{document}